\documentclass{article}
\usepackage{aaai16}
\usepackage{graphicx}
\usepackage[english]{babel}
\usepackage[normalem]{ulem}
\useunder{\uline}{\ul}{}
\usepackage{multirow}
\usepackage{booktabs}
\usepackage{caption}
\usepackage{amsmath}

\nocopyright{}

\begin{document}

\title{To Follow or Not to Follow: Analyzing the Growth Patterns of the Trumpists on Twitter\thanks{This paper will be presented at the first International Workshop on News and Public Opinion at ICWSM 2016 (May, Germany).}}

\author{Yu Wang\\Political Science\\University of Rochester\\Rochester, NY 14627\\ywang@ur.rochester.edu\And Jiebo Luo\\Computer Science\\University of Rochester\\Rochester, NY 14627\\jluo@cs.rochester.edu\And Richard Niemi\\Political Science\\University of Rochester\\Rochester, NY 14627\\niemi@rochester.edu \And Yuncheng Li\\Computer Science\\University of Rochester\\Rochester, NY 14627\\yli@cs.rochester.edu}
\maketitle

\begin{abstract} \small\baselineskip=9pt 


In this paper, we analyze the growth patterns of Donald Trump's followers (Trumpists, henceforth) on Twitter. We first construct a random walk model with a time trend to study the growth trend and the effects of public debates. We then analyze the relationship between Trump's activity on Twitter and the growth of his followers. Thirdly, we analyze the effects of such controversial events as calling for Muslim ban and his `schlonged' remark.

\end{abstract}

\section{Introduction} 
From proposing mass deportation of Mexican immigrants to calling for banning Muslims from entering the U.S. and more recently to saying that Hillary Clinton `got schlonged' by President Obama in 2008, Donald Trump has emerged as the most controversial candidate in the 2016 presidential race. Yet, for all the predictions of his fall from grace, accordingly to all the major polls, Trump is leading the GOP presidential race by a large margin.\footnote{For a detailed summary of the many poll results, please see http://elections.huffingtonpost.com/pollster/2016-national-gop-primary.}

\begin{figure}[h!]
\includegraphics[width=8.4cm]{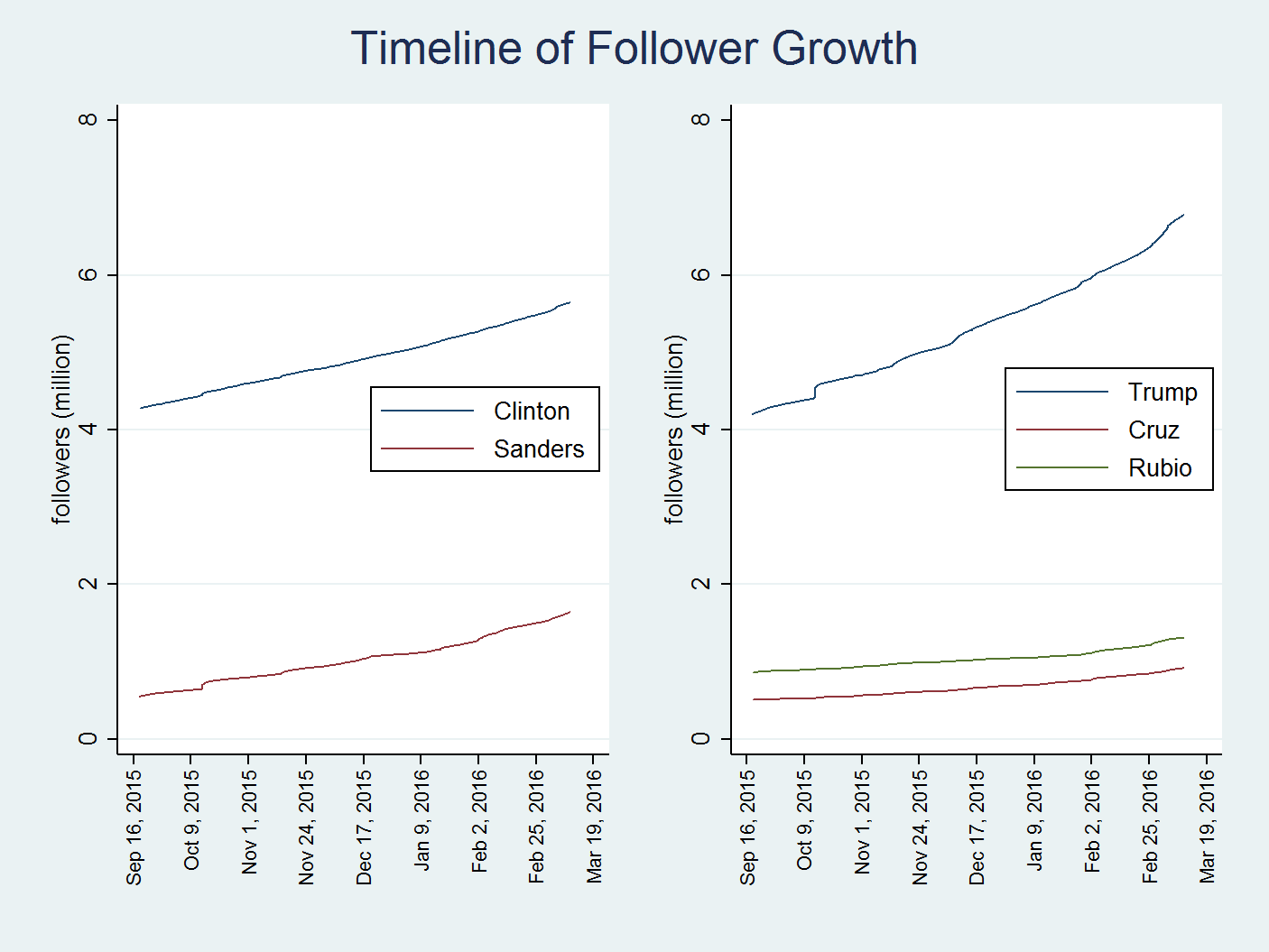}
\caption{Growth of Trump followers, compared with other major candidates}
\label{gallery}
\end{figure}

Trump's leading position also expands into the Twitter sphere, where he currently boasts of 6.58 million followers (Figure \ref{gallery}). By comparison, Ted Cruz, who enjoys the second highest Republican support, has 0.88 million followers and Marco Rubio, who regularly ranks third in poll numbers, has 1.28 million followers. We observe a large disparity in the number of followers between Donald Trump and other Republican candidates.\footnote{Data in our dataset \textit{US2016} shows that a gap of similar magnitude also exists in the Democratic presidential race, where Clinton ranks first, Sanders the second, and O'Malley the third.} In this work, we analyze how the political activities of Donald Trump affect public opinion and translate into growth dynamics of the Trumpists.

Our goal in this paper is threefold. We first compare the growth trend of the Trumpists with that of other candidates' followers. Along the way, we shall also evaluate the effects of the public debates on the growth rate of the Trumpists. Second, we study the relationship between Trump's tweeting patterns and the growth patterns of the Trumpists. Third, we quantitatively measure the effects of the above-cited controversies.

Our study shows that while Trump enjoys the strongest growth trend of all the major candidates, the growth itself is not helped by his performance in Republican debates. Our study also finds that the more tweets he posts, the faster the Trump camp grows. Lastly, we show that the above-cited two controversial events in Trump's campaign may have not turned public opinion against him.

\section{Related Work}

Our work builds upon previous research in both political science and computer science.

Political scientists have a long history of studying the effects of campaigns and public debates. Many studies have found that campaign and news media messages can alter voters' behavior \cite{politicalManipulation,newsmatters}. According to Gabriel S. Lenz, public debates help inform some of the voters about the parties' or candidates' positions on the important issues \cite{lenz}. In our work, we construct a random walk model with a time trend to estimate Trump's performance in the public debates. We also estimate how the intensity of Trump's tweeting activity affects the growth of the Trumpists.

There is a burgeoning literature in computer science on using social media data to analyze and predict elections. Research by \cite{tumasjan} finds that the number of messages mentioning a party reflects the election results. According to \cite{fbcount}, the number of Facebook fans constitutes an indicator of candidate viability. \cite{trumpists} use user profile images to study and compare the social demographics of Trump followers and Clinton followers. Similarly, our work is also motivated by the high parallel between performance in the polls and popularity in Twitter. \cite{trumponfire} employ LDA  to model tweet topics and use negative binomial regression on the number of tweet `likes' to infer topic preferences of Trump followers.



\section{Dataset}
We use the dataset $\textit{US2016}$, constructed by us with Twitter data. The dataset contains a tracking record of the number of followers for all the major candidates in the 2016 presidential race and is updated every ten minutes. It also contains the user names, geographical locations, number of tweets posted and the profile images of the followers for the major candidates. For our purpose, we will focus on the dynamics of Trump's followers. The dataset spans the whole period between September 18th, 2015 and December 22nd, 2015, contains 12,968 observations and covers three Democratic debates and three Republican debates. In addition, $\textit{US2016}$ also contains all the tweets (2113, in total) that Trump posted during the same period. We will use these tweets to learn Trump's tweeting pattern. Dates on which Trump-initiated controversies broke out are public knowledge and we obtain them online.

\section{Estimation Results}
In this section, we present our estimation results on Trump's debate performance, effects of his tweeting activity, and the effects of the above-cited controversies.

\subsection{Public Debates}
We assume that the number of followers follows a random walk and that there is a time trend that represents the strength of growth. This time trend will be affected by candidates' performance during the public debate and it might also deviate from the trend during weekends when there is decreased news media activity. Meanwhile, assuming that the act of following the candidates exhibits hourly patterns, e.g. relatively low during late night hours, we control for the hour of the day. Formally, we formulate the model as follows:
\begin{equation*} \label{eq1}
\begin{split}
\Delta \mathrm{Followers}_t &=\beta_0+\beta_1\Delta\mathrm{Time_t}+\beta_2\mathrm{Debate(D)}\\
 &+ \beta_3\mathrm{Debate(R)}+\beta_4\mathrm{Weekend}\\
 &+\pmb{\gamma}\cdot controls+\epsilon_t\\
\end{split}
\end{equation*}
where $\Delta$Followers$_{t}$ is the number of new followers at period t, $\Delta$Time$_t$ represents the time interval, Debate(D) is binary, denoting whether a Democratic debate is in effect, Debate(R) denotes whether a Republican debate is in effect, and Weekend is binary, taking value of 1 if the time is weekend and 0 otherwise, and lastly $controls$ are a set of hour dummies that control for the fact that Twitter activities are low during late night and early morning hours.

We then estimate the coefficient vector $\pmb{\beta}$ using OLS for Trump and five other candidates. $\beta_1$ represents the time trend: a larger $\beta_1$ represents faster growth. $\beta_2$ measures the effects of Democratic debates and $\beta_3$ measures the effects of Republican debates. For Trump, $\beta_3$ measures his debate performance and $\beta_2$ measures the effects of Democratic debates on the growth of his followers. Lastly, $\beta_4$ represents the effects of the weekend, when the intensity of activities on the news media is low.

We report the results in Table \ref{ols}. In terms of growth trend, Trump clearly leads the entire presidential race. In terms of debate performance, we find that both Carson and Rubio outperformed Trump in the Republican debates. We observe that Trump is winning lots of followers during Democratic debates.

Lastly, we find that during weekends the growth of Trumpists tends to slow down. This ``weekend'' effect is also significant for other candidates except Carson. We posit that this ``weekend'' effect might be caused by a decreased level of activity in the news media. This suggests that news media might be playing a significant role in the expansion of the Twitter sphere. Another possible reason is that the general public are less responsive to the news media during the weekend.

\begin{table*}[htbp]\centering
\def\sym#1{\ifmmode^{#1}\else\(^{#1}\)\fi}
\caption{Comparing Trump' Performance on Twitter with Other Candidates}
\label{ols}
\setlength{\tabcolsep}{13.5pt}
\begin{tabular}{l*{6}{c}}
\hline\hline
                    &\multicolumn{1}{c}{Clinton}&\multicolumn{1}{c}{Sanders}&\multicolumn{1}{c}{Biden}&\multicolumn{1}{c}{Trump}&\multicolumn{1}{c}{Carson}&\multicolumn{1}{c}{Rubio}\\
\hline
$\Delta$Time           &       67.50\sym{***}&       33.46\sym{***}&       10.61\sym{***}&       86.40\sym{***}&       35.54\sym{***}&       23.42\sym{***}\\
                    &     (0.239)         &     (0.499)         &    (0.0263)         &     (1.385)         &     (0.346)         &     (0.107)         \\
Democratic Debate           &       59.73\sym{***}&       140.9\sym{***}&       1.669\sym{***}&       235.0\sym{***}&       31.49\sym{***}&       16.17\sym{***}\\
                    &     (1.768)         &     (3.691)         &     (0.195)         &     (10.24)         &     (2.555)         &     (0.789)         \\
Republican Debate           &       8.109\sym{***}&       13.77\sym{***}&      -1.142\sym{***}&      -6.013         &       62.40\sym{***}&       29.73\sym{***}\\
                    &     (1.792)         &     (3.741)         &     (0.197)         &     (10.38)         &     (2.589)         &     (0.800)         \\
Weekend              &      -11.38\sym{***}&      -22.11\sym{***}&      -1.187\sym{***}&      -51.34\sym{***}&      -0.245         &      -3.208\sym{***}\\
                    &     (0.978)         &     (2.042)         &     (0.108)         &     (5.665)         &     (1.413)         &     (0.436)         \\
Hour Controls              &           Yes         &           Yes         &           Yes         &           Yes         &          Yes         &          Yes         \\
Constant            &       7.215\sym{***}&       21.58\sym{***}&      -1.329\sym{***}&       43.89\sym{***}&       10.90\sym{***}&      -3.031\sym{**} \\
                    &     (2.107)         &     (4.399)         &     (0.232)         &     (12.21)         &     (3.044)         &     (0.940)         \\
\hline
Observations        &       12971         &       12971         &       12971         &       12968         &       12969         &       12969         \\
Adjusted \(R^{2}\)  &       0.863         &       0.335         &       0.927         &       0.264         &       0.474         &       0.799         \\
\hline\hline
\multicolumn{7}{l}{\footnotesize Standard errors in parentheses}\\
\multicolumn{7}{l}{\footnotesize \sym{*} \(p<0.05\), \sym{**} \(p<0.01\), \sym{***} \(p<0.001\)}\\
\end{tabular}
\end{table*}

\subsection{Twitter Engagement}
\begin{figure}[h!]
\includegraphics[width=8.4cm]{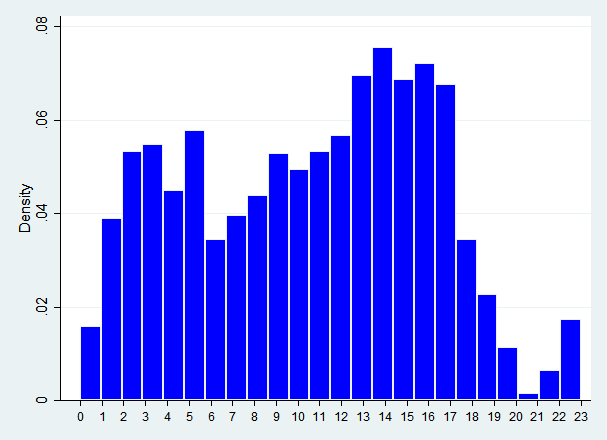}
\caption{The Daily Tweeting Pattern of Donald Trump}
\label{dailyPattern}
\end{figure}

\begin{figure}[h!]
\includegraphics[width=8.4cm]{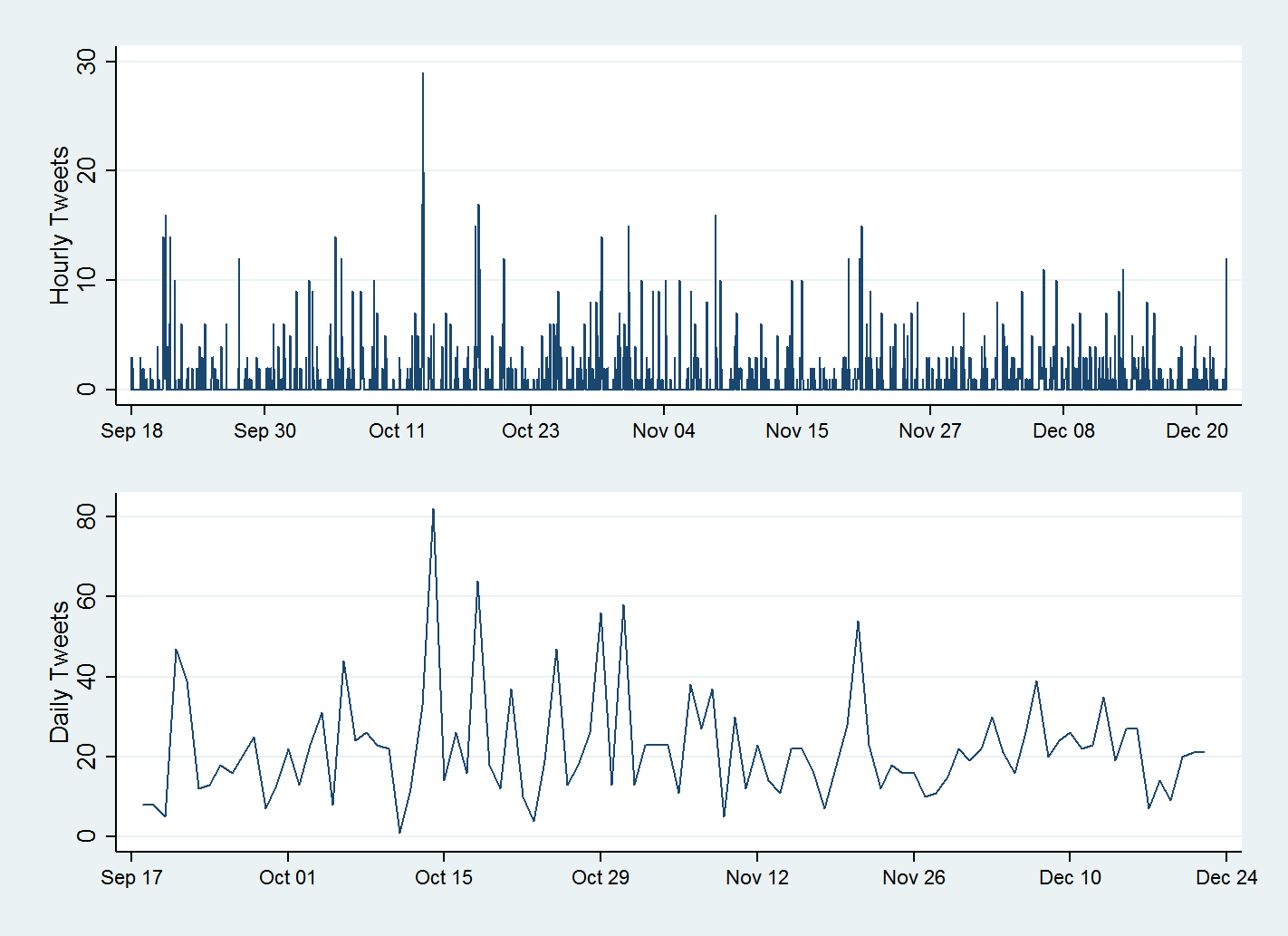}
\caption{Trump's Hourly and Daily Tweeting Intensity}
\label{dailyhourlyPattern}
\end{figure}

Motivated by the observation that Trump's tweeting activity exhibits a strong daily pattern, as shown in Figure \ref{dailyPattern}, and by the ``weekend'' effect uncovered from the first subsection, we study the relationship between Trump's tweeting intensity and the growth of Trumpists. To measure intensity, we calculate the number of tweets that Trump posted at each hour and then on each day during our observation period. Figure \ref{dailyhourlyPattern} presents our hourly and daily aggregation results. 

To measure the effects of Trump's tweeting intensity, we first test whether a rise in Trump's daily tweeting activity increases the growth of the Trumpists. Second, we test whether his hourly tweeting intensity increases Trumpist growth. Since there will be a lag effect between Trump posting a tweet and someone following Trump, here we include 8 lagged intensity values to test for Granger causality, following \cite{stockMarket}.

\begin{table}[]\centering
\def\sym#1{\ifmmode^{#1}\else\(^{#1}\)\fi}
\caption{Growth Patterns of the Trumpists}
\label{intensity}
\setlength{\tabcolsep}{4.5pt}
\renewcommand{\arraystretch}{1}
\begin{tabular}{l*{3}{c}}
\hline\hline
                    &\multicolumn{1}{c}{Baseline}&\multicolumn{1}{c}{Day}&\multicolumn{1}{c}{Hour}\\
\hline
$\Delta$Time           &       86.47\sym{***}&       86.51\sym{***}&       86.53\sym{***}\\
                    &     (1.395)         &     (1.403)         &     (1.398)         \\
Democratic Debate            &       234.3\sym{***}&       229.9\sym{***}&       233.1\sym{***}\\
                    &     (10.33)         &     (10.52)         &     (10.42)         \\
Republican Debate            &      -6.348         &      -11.73         &      -7.976         \\
                    &     (10.47)         &     (10.67)         &     (10.55)         \\
Weekend             &      -50.43\sym{***}&      -51.50\sym{***}&      -50.34\sym{***}\\
                    &     (5.713)         &     (5.834)         &     (5.733)         \\
Day (t)       &                     &       0.591\sym{**} &                     \\
                    &                     &     (0.194)         &                     \\
Hour(t)  &                     &                     &       0.614         \\
                    &                     &                     &     (1.224)         \\
Hour (t-1)  &                     &                     &     -0.0305         \\
                    &                     &                     &     (1.254)         \\
Hour (t-2)  &                     &                     &      -0.686         \\
                    &                     &                     &     (1.262)         \\
Hour (t-3) &                     &                     &      -1.578         \\
                    &                     &                     &     (1.249)         \\
Hour (t-4) &                     &                     &      -2.318         \\
                    &                     &                     &     (1.252)         \\
Hour (t-5)  &                     &                     &       3.188\sym{*}  \\
                    &                     &                     &     (1.256)         \\
Hour (t-6)  &                     &                     &       1.515         \\
                    &                     &                     &     (1.243)         \\
Hour (t-7)  &                     &                     &       2.735\sym{*}  \\
                    &                     &                     &     (1.254)         \\
Hour (t-8)  &                     &                     &      -0.507         \\
                    &                     &                     &     (1.218)         \\
Constant            &       36.71\sym{***}&       24.47\sym{***}&       34.06\sym{***}\\
                    &     (3.244)         &     (5.176)         &     (4.124)         \\
\hline
Observations        &       12968         &       12826         &       12914         \\
Adjusted \(R^{2}\)  &       0.251         &       0.251         &       0.252         \\
\hline\hline
\multicolumn{4}{l}{\footnotesize Standard errors in parentheses}\\
\multicolumn{4}{l}{\footnotesize \sym{*} \(p<0.05\), \sym{**} \(p<0.01\), \sym{***} \(p<0.001\)}\\
\end{tabular}
\end{table}
\newpage
The estimation framework is the same as in the subsection on public debates, except that we now add tweeting intensity as an explanatory variable. We present our results in Table \ref{intensity}. Column 1 serves as the baseline. Column 2 investigates the effects of Trump's daily tweeting intensity. Column 3 investigates the effects of his hourly tweeting activity. We find that the number of tweets that Trump posts per day increases the growth of the Trumpists and that such an increase is statistically significant. We also find that hourly tweeting activity will not have an effect until 5 hours later. We interpret this as the dissemination time interval.

\subsection{Effects of Controversial Remarks}
Trump has proved himself the most controversial candidate in the 2016 presidential race. Yet, despite all the controversies that he has surrounded himself with, at the time of writing, Trump is by far the front-runner in the GOP presidential race.

Our dataset $\textit{US2016}$ is ideally suited for analyzing the effects of these controversies on public opinion. For this purpose, we progressively introduce two dummy variables \textit{Muslim Ban Proposal} and \textit{`Schlong' Comment}, which take the value 1 after the respective occurrence of the event. \textit{Muslim Ban Proposal} is 1 from December 8th onward and \textit{`Schlong' Comment} is 1 on Dec 22nd. The estimation framework is the same as in the public debates subsection.\\

\begin{table}[h!]\centering
\def\sym#1{\ifmmode^{#1}\else\(^{#1}\)\fi}
\caption{Event Analysis}
\setlength{\tabcolsep}{4pt}
\renewcommand{\arraystretch}{0.92}
\label{controversy}
\begin{tabular}{l*{3}{c}}
\hline\hline
                    &\multicolumn{1}{c}{Baseline}&\multicolumn{1}{c}{\textit{Muslim}}&\multicolumn{1}{c}{\textit{Schlong}}\\
\hline
$\Delta$Time           &       86.36\sym{***}&       86.37\sym{***}&       86.38\sym{***}\\
                    &     (1.399)         &     (1.399)         &     (1.399)         \\
Democratic Debate           &       214.9\sym{***}&       209.6\sym{***}&       209.3\sym{***}\\
                    &     (10.12)         &     (10.21)         &     (10.22)         \\
Republican Debate            &       7.895         &       3.553         &       3.314         \\
                    &     (10.37)         &     (10.43)         &     (10.44)         \\
Muslim Ban Proposal          &                     &       26.26\sym{***}&       26.97\sym{***}\\
                    &                     &     (6.992)         &     (7.109)         \\
`Schlong' Comment             &                     &                     &      -19.03         \\
                    &                     &                     &     (34.29)         \\
Constant            &       22.53\sym{***}&       19.03\sym{***}&       19.06\sym{***}\\
                    &     (2.827)         &     (2.976)         &     (2.976)         \\
\hline
Observations        &       12968         &       12968         &       12968         \\
Adjusted \(R^{2}\)  &       0.247         &       0.247         &       0.247         \\
\hline\hline
\multicolumn{4}{l}{\footnotesize Standard errors in parentheses}\\
\multicolumn{4}{l}{\footnotesize \sym{*} \(p<0.05\), \sym{**} \(p<0.01\), \sym{***} \(p<0.001\)}\\
\end{tabular}
\end{table}

We present our results in Table \ref{controversy}. We find that the Muslim Ban effectively boosted the growth of the Trumpists. As a robustness check, we shorten the time window during which the $\textit{Muslim Ban Proposal}$ variable takes value of 1 to 10 days and 5 days progressively. The results we obtain remain positive and significant. By contrast, we do not find a positive boost from the `Schlong' comment. The estimated coefficient is negative, but it is not statistically significant.\\
\section{Conclusion and Future Work}
We have presented a study on the growth patterns of Trump followers on Twitter. We first constructed a random walk framework to model the growth patterns. Into the time trend, we added the public debate effects so that we were able to evaluate Trump's performance with regard to attracting followers. We then evaluated the effects of Trump's tweeting activity on the growth of his followers. We found that the more he tweeted the faster his follower camp grew. Lastly, we measured the effects of two Trump-initiated controversies. Based on our data, neither one is hurting his campaign.

We see great promise in our work and we believe that the rise of Donald Trump is a significant event in American politics. Our immediate next step is to understand the demographics of the growing Trumpists and evaluate their sentiments through, e.g., tweets.

\section{Acknowledgment}
We gratefully acknowledge support from the University of Rochester, New York State through the Goergen Institute for Data Science, and our corporate sponsors Xerox and Yahoo.

\bibliographystyle{aaai}
\bibliography{yu}
\end{document}